\documentclass[conference,letterpaper]{IEEEtran}
\usepackage{makeidx}
\usepackage[T1]{fontenc}\usepackage[utf8]{inputenc}
\usepackage{url}
\usepackage{graphicx}
\DeclareGraphicsExtensions{.pdf,.png,.jpg}
\graphicspath{{./img/}}

\usepackage{algpseudocode}
\usepackage{algorithm}
\usepackage{varwidth}

\usepackage{flushend}
\usepackage{amsmath}
\usepackage{amssymb}
\usepackage{glossaries}
\usepackage[super]{nth}
\usepackage{multirow}
\usepackage{textcomp}
\usepackage{wrapfig}
\usepackage[lofdepth,lotdepth]{subfig}
\usepackage{url}

\newcommand{\coe}{$CO_{2}e$}
\newcommand{\n}{\oldstylenums}

\newacronym{coe}{\coe}{$CO_{2}$ equivalent}
\newacronym{ewma}{EWMA}{exponentially weighted moving average}
\newacronym{dvfs}{DVFS}{dynamic voltage \& frequency scaling}
\newacronym{vm}{VM}{virtual machine}
\newacronym{era}{ERA}{energy reduction assets}
\newacronym{api}{API}{application programming interface}
\newacronym{os}{OS}{operating system}
\newacronym{rtp}{RTP}{real-time pricing}
\newacronym{qos}{QOS}{quality of service}
\newacronym{sla}{SLA}{service level agreement}
\newacronym{pue}{PUE}{power usage efficiency}
\newacronym{cue}{CUE}{carbon usage effectiveness}
\newacronym{cef}{CEF}{carbon emission factor}

\begin{document}

\title{Take a break: cloud scheduling optimized for real-time electricity pricing}

 \author{
 
 \IEEEauthorblockN{Dražen Lučanin}
 \IEEEauthorblockA{Vienna University of Technology\\
 Email: drazen@infosys.tuwien.ac.at}
 
 \and
 
 \IEEEauthorblockN{Ivona Brandic}
 \IEEEauthorblockA{Vienna University of Technology\\
 Email: ivona@infosys.tuwien.ac.at}
 
 }

\maketitle


\begin{abstract}
Cloud computing revolutionised the industry with its elastic, on-demand\
approach to computational resources, but\
has lead to a tremendous impact on the environment.\
Data centers constitute 1.1--1.5\% of total electricity usage in the world.\
Taking a more informed view of the electrical\
grid by analysing real-time electricity prices,\
we set the foundations of a grid-conscious cloud.\
We propose a scheduling algorithm that predicts\
electricity price peaks and throttles energy consumption\
by pausing virtual machines.\
We evaluate the approach on the\
OpenStack cloud manager through an empirical\
approach and show reductions in energy consumption and costs.
Finally, we define\
green instances in which cloud providers can offer\
such services to their customers under better\
pricing options.


\end{abstract}

\begin{keywords}
cloud computing, scheduling, energy efficiency, green instances, electricity price,
smart grids, virtualisation.
\end{keywords}

\section{Introduction}

Due to cloud computing's surge in popularity, data centers running\
thousands of servers are\
being built all over the world to satisfy user demand.\
Such infrastructures have a huge impact on the environment---they account\
for 1.1--1.5\% of total electricity usage \cite{koomey_worldwide_2008}\
and constitute\
the majority of ICT sector's 2\% of all the \gls{coe} emissions
\cite{_gartner_????-1}.\
Existing scheduling methods in cloud computing typically\
monitor usage and performance, adhering to the agreed constraints\
and improving energy efficiency in a best-effort manner by throttling\
any surplus resources \cite{maurer_enacting_2011,beloglazov_energy-aware_????}.\
What such methods usually neglect is the complexity\
of the electrical grid that powers data centers and that not all\
energy has the same environmental impact---it differs between\
generators, depends on demand, grid congestion, time of day, weather\
and many other factors.\
An important aspect of current electrical grids is that due to huge demand\
peaks during midday, electricity often gets generated in more environmentally\ 
unsustainable ways such as by\
diesel generators \cite{andrews_potential_2008}, which is reflected\
in higher real-time prices \cite{klingert_sustainable_2012}.


Our \textbf{research question} is---\emph{how can scheduling in cloud computing be\
optimized for real-time electricity pricing to improve energy and cost\
efficiency?} To answer this challenge we introduce the grid-conscious cloud model\
which relies on\
computation elasticity to optimise energy usage under dynamic electricity prices.\
Building on this principle, we present a scheduling method\ 
that analyses historical electricity price data \cite{_ameren_????}\
to determine the most probable peak hours\
and then pauses the managed \gls{vm} during this period.\
We implemented this scheduling method---the peak pauser\footnote{\
Code and datasets:\
\url{launchpad.net/philharmonic} and \url{launchpad.net/berserk}},\
on the industry-standard OpenStack\
cloud manager \cite{ken_deploying_2011} to be able\
to empirically measure its effectiveness.\
We calculated further savings projections\
based on the available assessments of hardware\
used in Google's production environment \cite{fan_power_2007}\
and obtained encouraging results.\
Since pausing \gls{vm}s is an invasive action\
towards the user, we present \emph{green instances}, a business option\
similar to Amazon's spot instances \cite{agmon_ben-yehuda_deconstructing_2011}\
where the user is offered\
a better service price in exchange for reduced \gls{vm} availability.

Our \textbf{key contributions} are:\
(1) the grid-conscious cloud model that integrates cloud computing\
into the dynamically changing landscape of the electrical grid; 
(2) the peak pauser scheduling algorithm, shown to improve\
energy and cost efficiency in a realistic cloud environment;\
(3) green instances as a justification of reduced\
availability for customers as an opt-in, better pricing option that\
encourages environmentally-friendly behaviour.


\begin{figure*}
\includegraphics[width=0.95\textwidth]{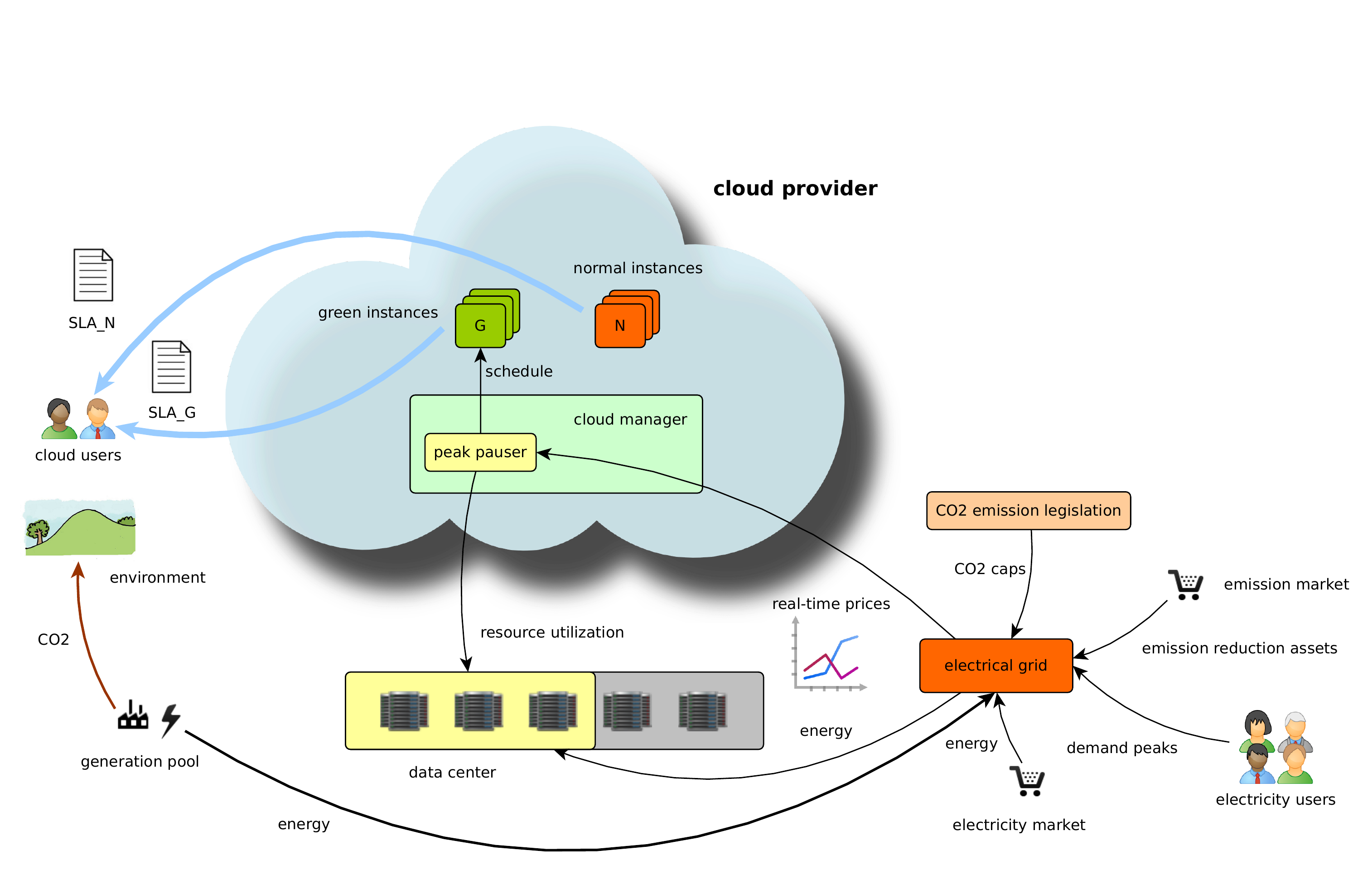}
\caption{The grid-conscious cloud model.}
\label{fig:grid-conscious_cloud}
\end{figure*}

\section{Related work}
\label{sec:related}
%
An analysis of historical electricity prices\
is given in \cite{huisman_history_2013}, showing that it is a very\
volatile signal, but exhibiting seasonality and regular peaks.\
Electricity price characteristics and day-ahead price calculation\
details are described in \cite{eydeland_energy_2002}.\ 
More generally, the act of using information provided by electrical grids\
to optimize usage efficiency falls in the domain of smart grids \cite{hassan_survey_2010}.\

There exists a large body of work regarding scheduling in cloud computing.\
Energy efficiency is often considered in cloud scheduling (a\
recent survey is available in \cite{berl_energy-efficient_2010}),\
although often with an overly simplistic view of the electrical grid---see\
e.g. \cite{maurer_enacting_2011,da_costa_green-net_2009,
beloglazov_energy-aware_????,wang_energy-efficient_2011,maurer_simulating_2010}.\
Insights about choosing hardware based on power consumption\
are given in \cite{beloglazov_energy-aware_????, beloglazov_energy_2010}.\
A more market-based system based on bidding and adapting to user needs\
is discussed in \cite{chase_managing_2001}.\
Unlike these studies, we consider the information provided through\
real-time electricity pricing for a more informed approach.\
In recent studies \cite{fard_multi-objective_2012,yu_multi-objective_2007},\
multi-objective approaches are becoming popular,\
where it is seen that the scheduler's goal is to satisfy economic cost,\
energy consumption and reliability constraints, which is nearer to our\
approach of weighing service availability and energy cost and efficiency.
On a more theoretical side,\
Garg et al. discussed cloud computing in the context of sustainability in\
\cite{garg_environment-conscious_2011}, with an emphasis on \coe\ emissions.\
We expanded on this in the context of the Kyoto protocol\
and its cap-and-trade system\
in our previous research \cite{lucanin_energy_2012}, which can be seen\
as a more general frame for the analysis of energy sources in cloud\
computing.


There exist several attempts to optimize large-scale computing for electricity prices.\
A theoretical analysis of scheduling with regard to electricity prices, job\
queue lengths and server availability is given in \cite{ren_provably-efficient_2012},\
but no statistical seasonality of electricity prices was assumed (which does exist as\
shown in \cite{huisman_history_2013}) to improve scheduling decisions.\
Qureshi et al. simulate potential gains from temporally- and geographically-aware\
content delivery networks in \cite{qureshi_cutting_2009}, with predicted electricity cost\
savings of up to 45\%. This shows that there indeed is potential in pairing distributed computing management\
with knowledge about the electrical grid.\
Aikema et al. \cite{aikema_energy-cost-aware_2011} show a method of rescheduling low-priority\
jobs in clusters to improve the data center's\
carbon footprint, which goes in the direction\
of our green instances in the context of computational grids.\
A similar approach, only for migration geographically-distributed data centers\
is presented by Buchbinder et al. in \cite{buchbinder_online_2011}.

\section{Foundations of the grid-conscious cloud}



The grid-conscious cloud model is illustrated in\
Fig \ref{fig:grid-conscious_cloud}.\
Assuming a real-time electricity pricing model,\
the price of electricity offered by the utility changes hourly.\
Various factors such as active generation pools, electricity markets,\
volatile demand \cite{weber_uncertainty_2004,weron_modeling_2006},\
\gls{coe} emission legislation and markets \cite{lucanin_energy_2012} that\
influence the electrical grid are condensed in this price signal.\
Combined with the elasticity of cloud computing, this creates an\
opportunity to optimize the conversion of energy to computation\
that would benefit both sides.\
%

The grid-conscious cloud is based on the idea that \emph{computation is more\
elastic than energy.} Moving energy around, storing it or building infrastructure\
to satisfy demand can be very expensive and environmentally inefficient\
\cite{weber_uncertainty_2004}.\
Computation, on the other hand---especially using modern paradigms such as\
virtualisation and cloud computing, can be scaled outwards or inwards,\
moved to other locations and postponed for a later time, depending on the\
needs of users and resource providers.

The provider of a grid-conscious cloud is\
motivated to weigh the electricity price parameter in its scheduling and\
resource utilization process. It needs a scheduler\
that can respond to changing electricity prices,\
altering the cloud's physical resource usage and energy demand.\
Changing resource usage potentially affects end users so it\
might be necessary to open a new business interface towards the end users.\ 
Taking all this into account, the grid-conscious cloud needs\
to satisfy three major requirements which we examine in the following subsections:\
(1) a real-time pricing option of buying electricity, (2) a resource scheduler\
that considers electricity prices to throttle energy demand when\
they are high and (3) a business model to offer this kind of service to end users\
with a justification for potential performance degradation.

\subsection{Background: real-time electricity prices}
\label{sec:electricity}


\begin{figure*}
\centering
\subfloat[][Hourly prices---single day and average]{
\includegraphics[width=0.5\textwidth]{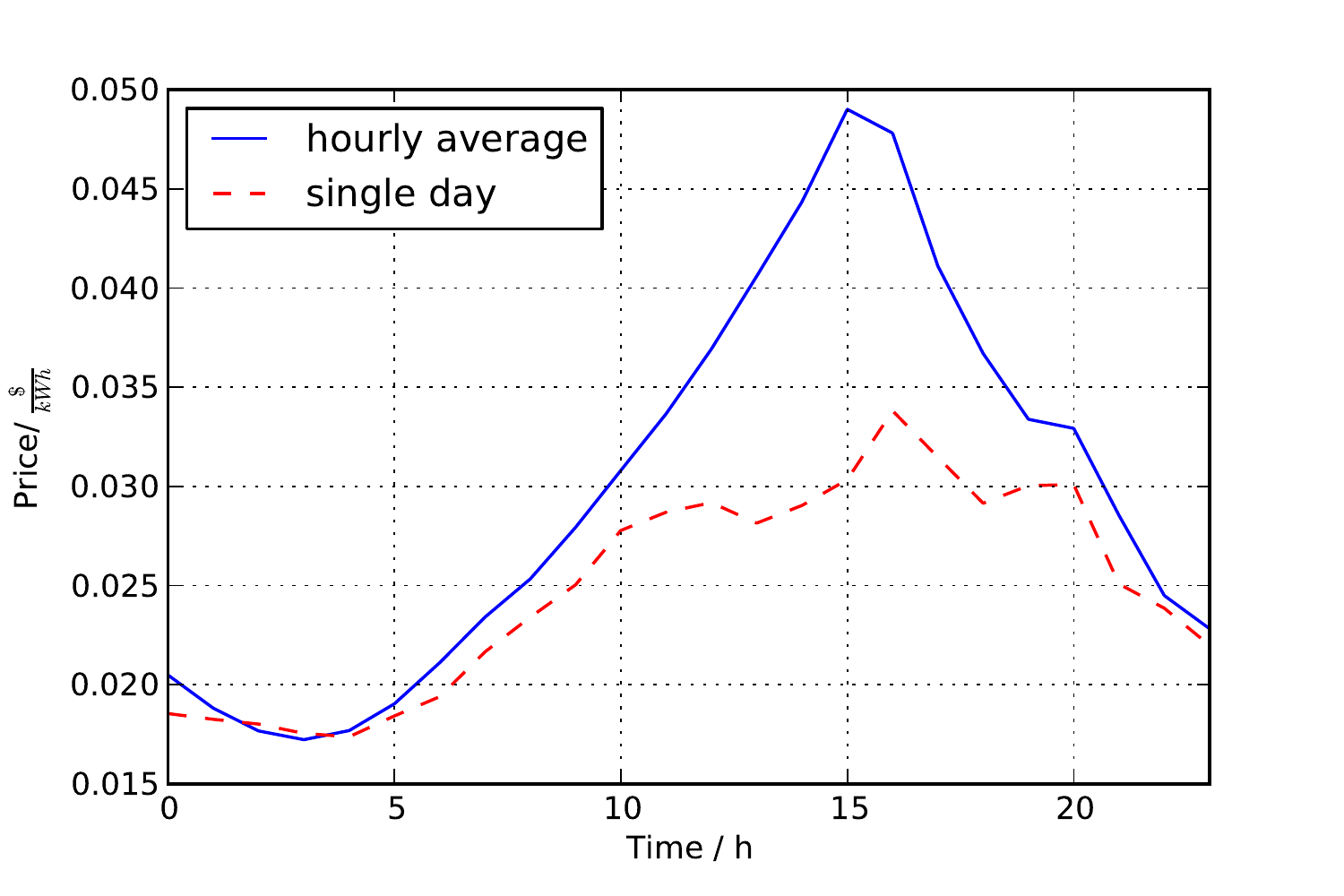}
\label{fig:prices-avg-single}
}
\subfloat[][Peak hour distribution]{
\includegraphics[width=0.5\textwidth]{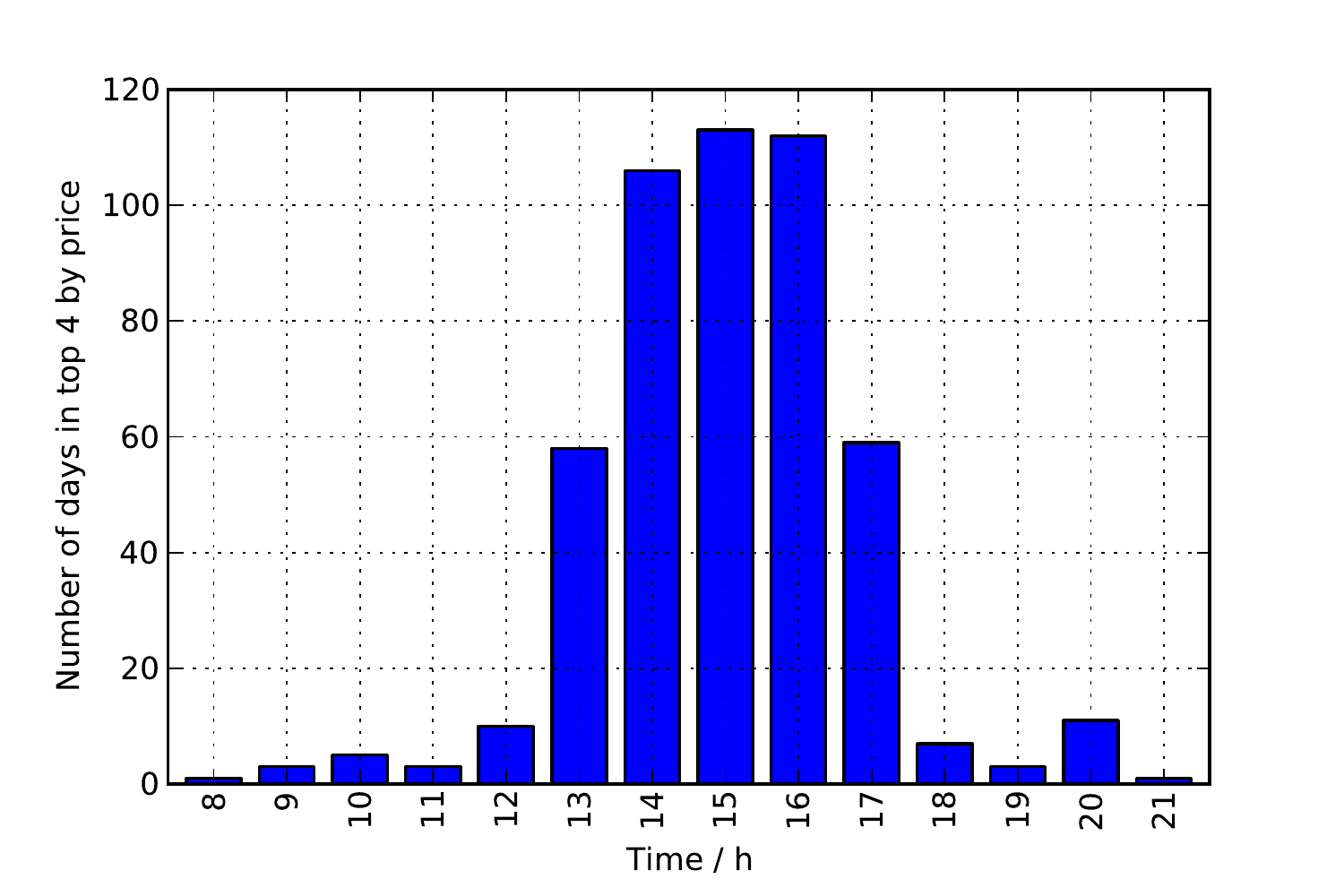}
\label{fig:expensive_histogram}
}
\caption{Historical electricity prices (data taken from \cite{_ameren_????})}
\end{figure*}

Peak-load pricing is a technique where demands for a public good\
in different periods of the day, month or year are considered\
to find the optimal capacity and the optimal peak-load prices.\
The method was being developed\
as far back as 1949 \cite{boiteux_peak-load_1960}.\
This is the basis of electricity spot markets,\
where electricity can be bought\
at real-time prices \cite{schweppe_spot_1988}.

Many utilities (such as Ameren \cite{_ameren_????}) offer\
a \gls{rtp} option\
where electricity prices change hourly,\
based on market supply and demand. Generally speaking,\
market prices are highest during times of peak demand\
(during daytime, usually peaking at 15:00), as\
shown in Fig. \ref{fig:prices-avg-single}. In \gls{rtp}s,\
customers are given price signals\
to guide their\
energy use and ease the burden on utilities during peak hours.\
Such customers are rewarded by potentially saving money\
when compared to the standard rate.\
There exist other pricing options, such as \emph{critical peak pricing}\
and \emph{time of use pricing} which could be suitable for\
our grid-conscious cloud model. In this study, however, we focus only\
on \gls{rtp}---considered to be the most direct and efficient\
demand-response program \cite{albadi_demand_2007}.   

An important point to consider that was analysed in\
\cite{klingert_sustainable_2012}\
is that in some cases utility companies have to resort to\
environmentally more harmful energy generation methods to handle peak demand.\
This is true in the UK, for example, where the National Grid can call\
upon about 2 GW of diesel-generated power to meet demand spikes,\
resulting in hundreds of hours of expensive and harmful fuel usage\
annually \cite{andrews_potential_2008}.\
The result is that a high electricity price also\
indicates a high rate of \gls{coe} emissions.\
This proportionality is further fortified under the Kyoto protocol's\
cap-and-trade model, where\
the \coe\ emissions themselves carry a market price as we analysed in our\
previous work \cite{lucanin_energy_2012}. This additional \coe\ emission price\
increases the price of energy coming from environmentally harmful sources\
even more.\
These specifics vary between different locations, however, which underlines\
the importance of having access to information grid through\
smart grids to make more informed, efficient decisions.

Another viewpoint of real-time prices is that it encourages polite behaviour,\
complying with the utilities' needs. This enables\
utilities to be more efficient (in terms of cost efficiency, reliability and\
environmental impact) and is said to create \emph{energy\
reduction assets} \cite{_understanding_????}.\
Such assets can then be exchanged for certain other\
benefits, such as lower energy prices, meaning that they\
benefit both sides.\
These power-reduction assets can be achieved more directly by integrating\
our grid-conscious cloud model with a direct demand/response control\
mechanism available in modern smart grids \cite{hassan_survey_2010}.

These examples show that there exist both monetary and ecological\
incentives for reducing energy consumption during hours of electricity\
price peaks---especially in computing clouds, which are substantial\
energy consumers \cite{koomey_worldwide_2008}.


%
%

\subsection{Optimizing the scheduler: the peak pauser algorithm}
\label{sec:scheduler}

The aim of this algorithm is to curtail energy consumption of a cloud\
during the hours of the day with the highest electricity price.\
Its pseudo-code is shown in Alg. \ref{alg:pp}.

The algorithm first\
finds $n$ hours in a day that are statistically most probable to be\
the peak-price hours.\
This number is defined using the parameter $downtime\_ratio$ given to the algorithm\
and is defined as
\
\begin{equation}
downtime\_ratio = \frac{\text{n}}{24}
\end{equation}
\
We then define the set of \emph{expensive hours} by calculating\
hourly price averages over the provided historical dataset, sorting them descending\
and selecting the first $n$ hours in the $find\_expensive\_hours$ function\
that requires historical electricity price data such as \cite{_ameren_????}.

%

We verified this function by analysing how often certain hours of the\
day appear among the top 4 by price (Fig. \ref{fig:expensive_histogram}\
shows the regular cyclic nature with peaks in the afternoon)\
and found it to result in only a negligible error\
\footnote{The root-mean-square error of the sum of expensive hours in a day\
determined by our function compared to an optimal daily-changing set\
that assumes a priori price knowledge\
equals $0.0058\ \$/kWh$ or  $\approx 3\% $ of the absolute amounts.}
compared to an\
ideal scenario where we know the prices in advance, sufficiently good\
for the purposes of our discussion.

Knowing how to determine expensive hours, we can define\
the scheduling algorithm as an endless loop\
(the $peak\_pauser$ function) that checks if the electricity is predicted\
to be expensive at the moment (in the $is\_expensive$ function that\
returns a boolean value depending on the current time's membership in the\
$expensive\_hours$ set).\
If the price is expensive, the scheduler pauses the set $G$ of instances\
it controls (green instances, which we will further discuss in the next section)\
and if it is not expensive, instances in $G$ are unpaused (or left running if\
they were not paused before). The scheduler can then remain idle for the remainder\
of the hour, so as not to waste resources.

\begin{algorithm}
\caption{The peak pauser algorithm.}
\label{alg:pp}
\begin{algorithmic}

\Function{find\_expensive\_hours}{$downtime\_ratio$}
\State $prices\gets\text{historical hourly prices}$
\State $avg\_prices\gets$ group $prices$ by hour and calculate mean
\State sort $avg\_prices$ descending by price
\State $n\gets ceil(downtime\_ratio*24)$\Comment{$ceil$: find first larger integer}
\State $expensive\_hours\gets$ {first $n$ elements of $avg\_prices$}
\State \Return $expensive\_hours$
\EndFunction
\item[]

\State  \begin{varwidth}[t]{\linewidth}
	$expensive\_hours\gets find\_expensive\_hours($\par
		\hskip\algorithmicindent $downtime\_ratio)$
\end{varwidth}

\item[]
\Function{is\_expensive}{}
	\State $time\gets\text{current time of day}$
	\State\Return $time.hour\in expensive\_hours$
\EndFunction
\item[]

\Function{peak\_pauser}{$G$}
\While{$True$}
	\If{$is\_expensive()$}
		\State pause $\forall instance \in G$
	\Else
		\State unpause $\forall \text{paused instance} \in G$
	\EndIf
	\State idle for the remainder of the hour
\EndWhile
\EndFunction

\end{algorithmic}
\end{algorithm}

A possible alternative to pausing would be to switch to\
battery power supply \cite{palasamudram_using_2012,bianchini_parasol:_????}\
during expensive hours.\
Additional logic can be added to the algorithm by dynamically\
determining duration of the pause interval (the parameter $downtime\_ratio$),\
based on e.g. considering the\
current day's deviation from monthly or annual averages. This\
way we could have longer pause periods during unusually ``expensive'' days\
and close-to-normal operation on ``cheaper'' days. For the purposes\
of our evaluation, we chose a predefined value $downtime\_ratio=0.16$,\
yielding in 4 paused hours.

\subsection{The green instance model}

To justify \gls{vm} pausing to users, we propose \emph{green instances}\
as an option\
in the manner of Amazon's \emph{spot instances}\
\cite{agmon_ben-yehuda_deconstructing_2011}.\
Spot instances\
have a disadvantage---they will only be running when\
there are free computing resources; and an advantage---a\
more affordable price. Expanding on this philosophy of acceptable\
trade-offs, we devised the green instance model.\
We envision it as an option where\
(1) compute instances are offered at a reduced availability time,\
but (2) at a lower price and (3) with environmental metrics\
presented to the user.\
From the cloud provider's perspective,\
this allows for more flexibility\
while scheduling to reduce energy costs and lessen the environmental\
impact which is good for public relations.

This would be an opt-in model, where users would be offered an additional\
\gls{sla}---$SLA_G$ to rent green instances or choose normal instances\
($SLA_N$). The set $G$ of instances managed\
by the peak pauser would be restricted to green instances only.\

Only users who applied for green instances and\
therefore accept occasional downtimes (at relatively\
predictable times of day) would feel the consequences of the peak pauser\
and create\
\gls{era}s. \gls{era}s basically mean that the cloud provider\
gets better electricity prices in return for\
helping the electrical grid's efficient operation.\
As a result, the cloud provider\
can offer lower prices to the end user, to compensate for the reduced\
\gls{vm} availability. So, where in spot instances\
users sacrifice performance for profitability, in green instances they\
sacrifice availability for profitability.


As a further reinforcing factor in favour of green instances,\
users could be presented with the approximated energy savings arising\
from choosing green over normal instances.\
Curry et al. \cite{curry_environmental_2012} denote this type of information\
\emph{environmental charge-backs} ($EC$). Using \gls{cef} and \gls{pue}\
defined in \cite{belady_carbon_2010} we could express it as:
\
\begin{equation}
EC = CEF * PUE * (\text{VM energy consumption})
\end{equation}
\
VMeter developed in \cite{bohra_vmeter:_2010} might be used to determine\
the energy consumption of a user's \gls{vm}s or the approach from\
\cite{alonso_tools_2012} for parallel applications.

Green instances would target the same user group as Amazon's spot instances do\
and the sole fact that Amazon, a commercial corporation, still offers this\
service shows that there are interested parties.\
As a potential use-case,\
many automated background processes would be fit for running on green instances.\
Examples of such processes are nightly builds of software (frequent\
and long-lasting due to compile-time optimization for various architectures),\
automated testing, web crawling, offline data mining etc.\
However, for applications that require\
real-time human interaction, access to data in a storage-as-a-service manner\
\cite{livenson_towards_2011}, normal instances would be preferred.


\section{Evaluation methodology}
\label{sec:methodology}

The goal of the empirical testing environment is to show how the peak pauser\
scheduler influences energy efficiency and application execution time. \ 
To measure the effects of applying the peak pauser scheduler in practice,\
it was implemented on top of the open source cloud manager OpenStack \cite{ken_deploying_2011},\
hosting a \gls{vm}.\
The server running the experiment was connected to a wattmeter that provides\
us with precise information about power consumption.\
Additionally, to evaluate our scheduler under more realistic conditions than our\
prototype environment, we simulated the scheduler's effects\
in production environments, such as that of Google \cite{fan_power_2007}.   

We will now go through the details of our evaluation setup and show how\
we measured the effectiveness of the peak pauser algorithm empirically and\
by simulating production systems.

\subsection{Empirical testing}

\begin{figure}
\includegraphics[width=0.98\columnwidth]{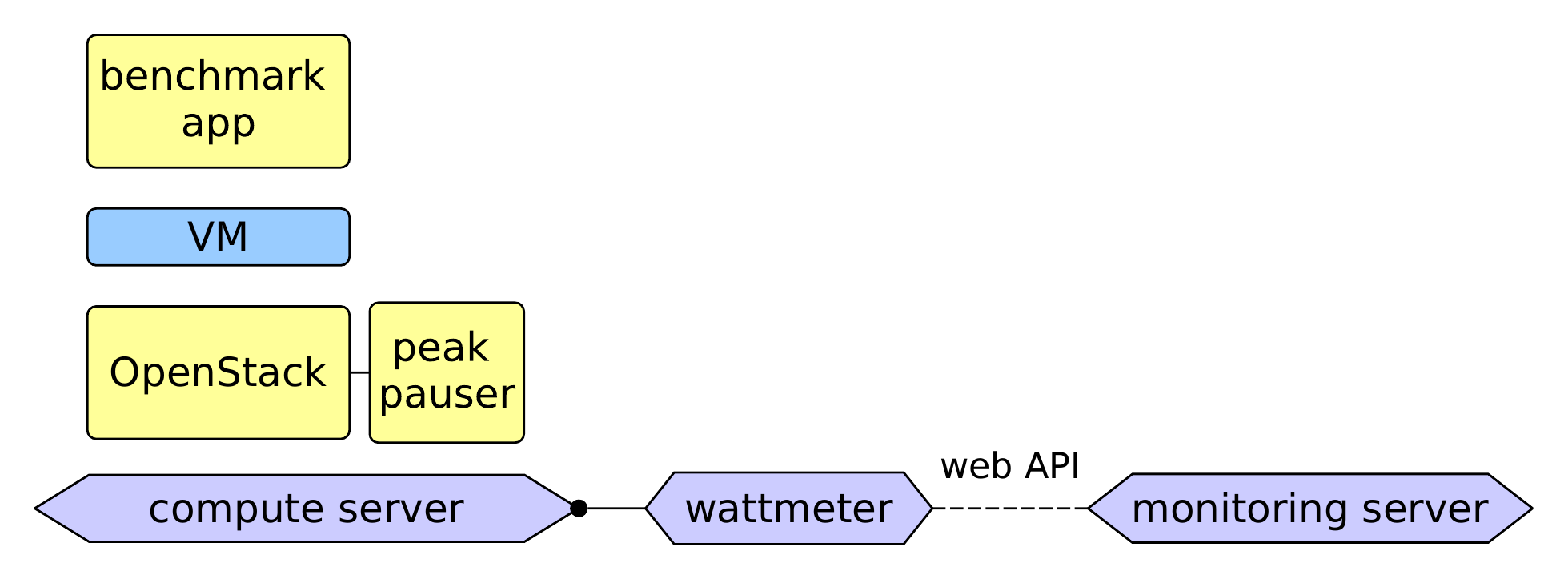}
\caption{Experiment deployment}
\label{fig:deployment}
\end{figure}

We ran a predictable\
synthetic benchmark application inside a \gls{vm} controlled by the peak pauser\
for approximately \n{24} hours.\
Exactly the same experiment was performed once again, but\
\emph{without a scheduler} to obtain comparison results.\
Two assumptions were made due to electrical price data availability---that\
the data center is located in Illinois, USA\
which offers a real-time pricing option \cite{_ameren_????} and that\
the experiment occurred a couple of weeks in the past.

The parameters of the peak pauser were set to pausing for a total\
of 4 hours in a day ($downtime\_ratio=0.16$). The statistically most probable peak hours\
were determined according to 3 months of historical electricity prices\
\cite{_ameren_????} before (non-inclusive)\
the day the experiment was assumed to be running on. 

As a benchmark application we used a CPU-intensive task---repeated\
recursive calculation of Fibonacci numbers.\
The benchmark application was run inside a \gls{vm}\
deployed on an OpenStack \cite{ken_deploying_2011} compute server\
\footnote{AMD Opteron 4130 2.6 GHz CPU, 8 GB RAM, Ubuntu 12.04\
server 64-bit GNU/Linux OS with OpenStack Essex}\
whose power consumption was measured\
%
%
as illustrated in Fig. \ref{fig:deployment}.\
The peak pauser scheduler was run on the same physical server as OpenStack and\
used its API to control the \gls{vm}.\



We measured \emph{active power} consumed by the compute server using a wattmeter\
\footnote{EATON ePDU PW104MA0UC34}\
offering a web interface for collecting data.\

\subsection{Estimating savings in production systems}
\label{sec:synthetic}

To broaden our evaluation by\
estimating savings in\
today's production systems, we relied on a study by Google\
\cite{fan_power_2007} to gain insight into their power consumption during peak and low demand.\
According to their study, peak power consumption of a\
server ranges from 100 to 250 W and\
idle power (the power consumed when nothing is executed on this server) can be\
as low as 50-65\% of this amount. This ratio of idle and peak power is referred to\
as the \emph{idle ratio}. It represents the energy elasticity of a server and\
turned out to be very important in our study.\

In addition to this, existing mechanisms such as suspend and wake-on-lan are\
already available for completely turning off under-utilized\
components. In fact the topic of energy-proportional servers is\
currently being studied extensively\
\cite{meisner_powernap:_2009}\
and it is likely that energy elasticity will\
improve in the future, aiming for an ideal idle ratio of 0.

\begin{figure}
\includegraphics[width=0.95\columnwidth]{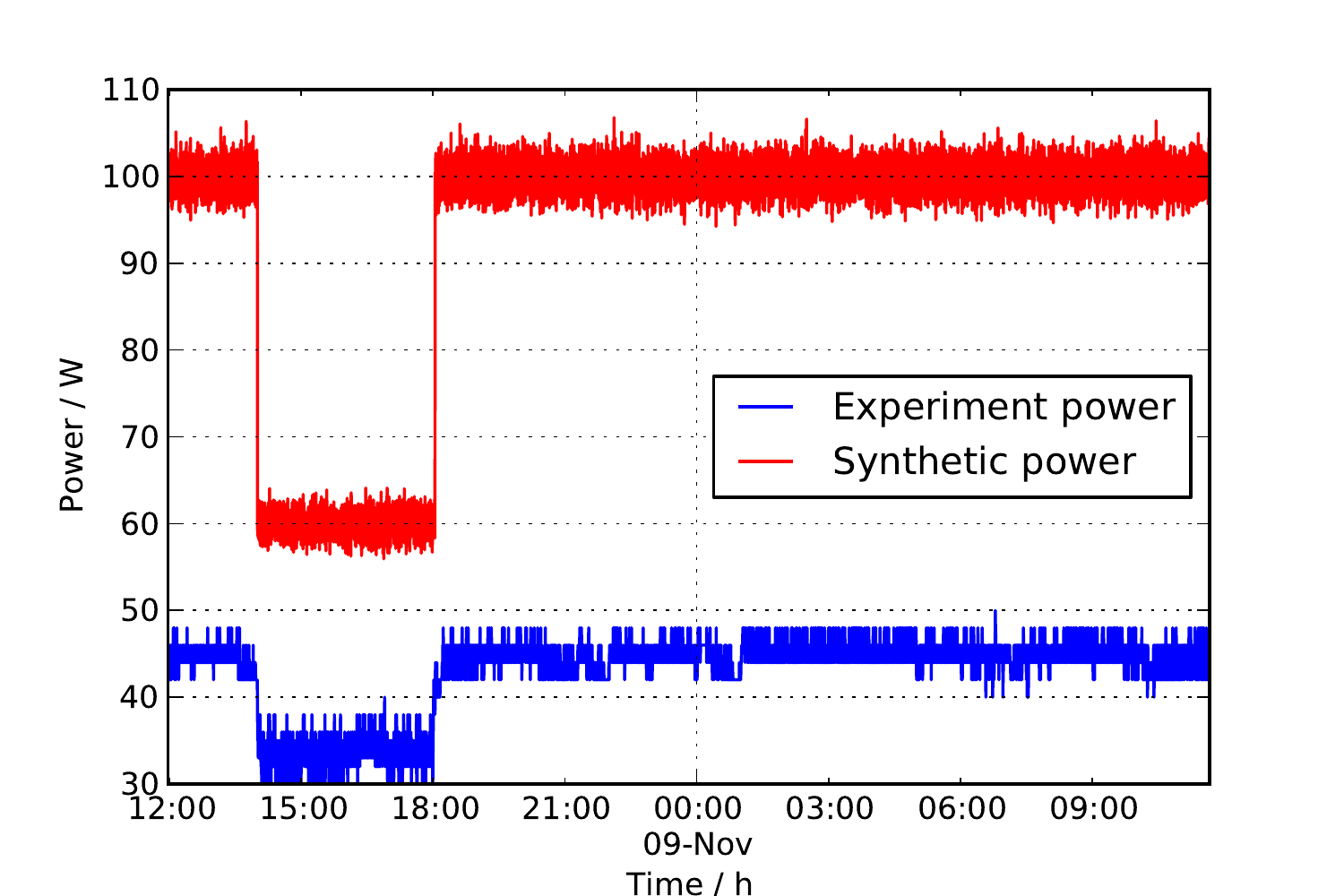}
\caption{A comparison of power consumption in our experiment (with peak pauser management)\
and a derived synthetic signal centered around 100 W peak power during \gls{vm} execution\
and 60 W idle power while the \gls{vm} is paused. The synthetic signal was used for estimating\
savings under different energy elasticity parameters.}
\label{fig:synthetic}
\end{figure}

A synthetic power time series was generated\
according to a simple model of\
our empirically collected data and scaled to match production-quality\
parameters from \cite{fan_power_2007}.\
We assumed normally distributed\
oscillation around the peak (during \gls{vm} execution) and idle\
(during a pause event) power values with a variance matching our experiment.\
The empirical and synthetic power signals\
can be seen in Fig. \ref{fig:synthetic}.

\subsection{Calculating electricity costs}

We previously established electricity costs to be more proportional to the\
actual environmental impact than the bare energy consumption,\
which we measure. To get as close as we can to knowing the\
impact our system has on the environment, we therefore\
want to calculate the total monetary expense. As we only have\
access to real low-level metrics obtained using the wattmeter\
to monitor server behaviour in an empirical experiment, we
have to calculate these monetary expenses on our own. To achieve\
this, we tried to mimic the pricing system actually used by the utilities\
in a real-time electricity price business model and we will explain\
these methods in detail here.

We sampled active power measurements of the compute server every 5 seconds.\
From this, we can calculate\
energy consumption and, according to real-time\
electricity prices \cite{_ameren_????} corresponding to\
the appropriate time intervals, derive a total energy price. $S_{total}$,\
the total electricity price in a time interval $T=\overline{t_0t_N}$\
is calculated from a numerical integral (we used the basic rectangle rule):

\begin{equation}
S_{total} =\sum_{t=t_0}^{t_N-1} \frac{t_N-t_0}{N}*P_t*C_t
\end{equation}

where $P_t$ and $C_t$ stand for power and electricity price\
in moment $t$, respectively and $N$ is the number of samples.

\section{Results}
\label{sec:results}

We will now present the results obtained after running the empirical\
experiment, followed by our estimation\
of savings in a production environment based on a synthetic power signal.\
Finally, we assemble the results into a green instance \gls{sla}\
that could be offered to users.

\subsection{Empirical savings}

\begin{figure*}
\centering
\subfloat[][Dynamic behaviour]{
\includegraphics[width=0.5\textwidth]{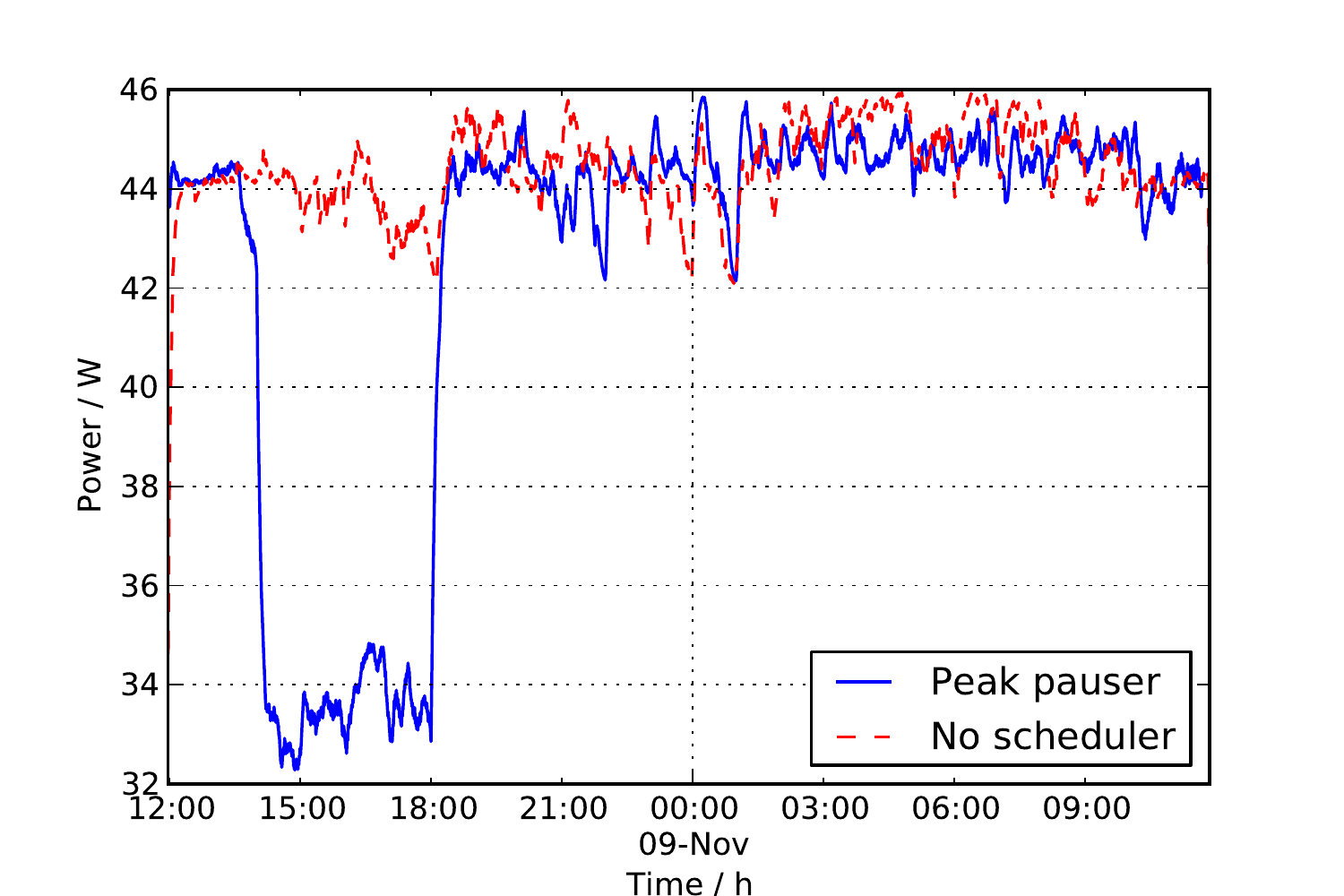}
\label{fig:dynamic_comparison}
}
\subfloat[][Aggregated results]{
\includegraphics[width=0.5\textwidth]{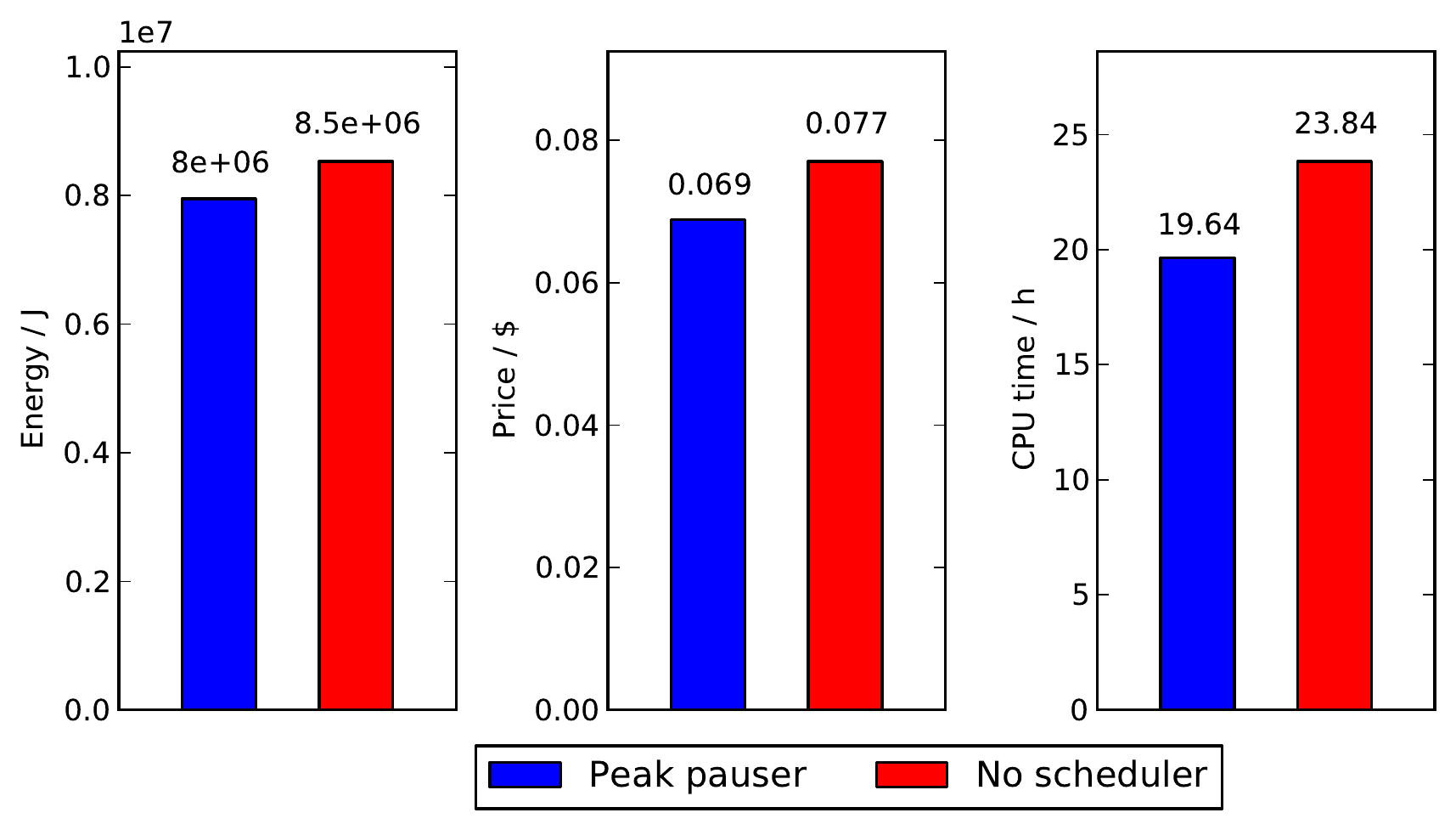}
\label{fig:aggregated_comparison}
}
\caption{Empirical evaluation results}
\end{figure*}

We are interested in a comparison of running the experiment for \n{24} hours\
with and without the peak pauser scheduler with regards to:
\begin{itemize}
  \item total consumed energy (also considering the energy consumed\
  while the \gls{vm} is paused)
  \item total electricity price based on dynamic, hourly charging
  \item benchmark CPU time
\end{itemize}

Runtime results can be seen in Fig. \ref{fig:dynamic_comparison}. The\
curves are smoothed using\
the \gls{ewma} method \cite{roberts_control_1959}. They\
show how the compute server's power consumption changes throughout\
the whole experiment. The drop in power from cca. \n{44} to {34} W\
($\approx 23\%$ or an idle ratio of $\approx 77\%$) can clearly be\
seen from \n{14} to \n{18} h. The \gls{vm}\
is paused during this time interval.

The results of aggregating measurements\
over the entire \n{24} hours of the experiment\
are shown in Fig. \ref{fig:aggregated_comparison}.\
The total energy consumption is $\approx 5.3\%$ lower\
than the amount consumed without\
a scheduler, due to the the reduced power consumption during \gls{vm} pausing.\
The difference in the electricity price is even\
larger, since the peak pauser only excludes the most expensive\
(financially and environmentally) hours of the day, the amount spent\
during a single day is $\approx 6.9\%$ lower when using the peak pauser.

The CPU time the benchmark application received in the experiment is\
4 hours less when using the peak pauser scheduler. This\
amounts to $\approx 17.6\%$ fewer actual calculations. This is a considerable\
performance deterioration,\
however it would only affect green instances whose owners agreed on\
fewer computing resources to increase energy efficiency and get\
a better price. An important thing to note here is\
that the current benchmark implementation\
only works in a single thread, not being able to consume all the CPU cores.\
In reality, where many processes are running, consuming many CPU cores,\
the ratio of energy and price savings to CPU time wastage would be\
greater, which we examine next.

\subsection{Projected savings}
\label{sec:results-projected}

\begin{figure}
\includegraphics[width=0.98\columnwidth]{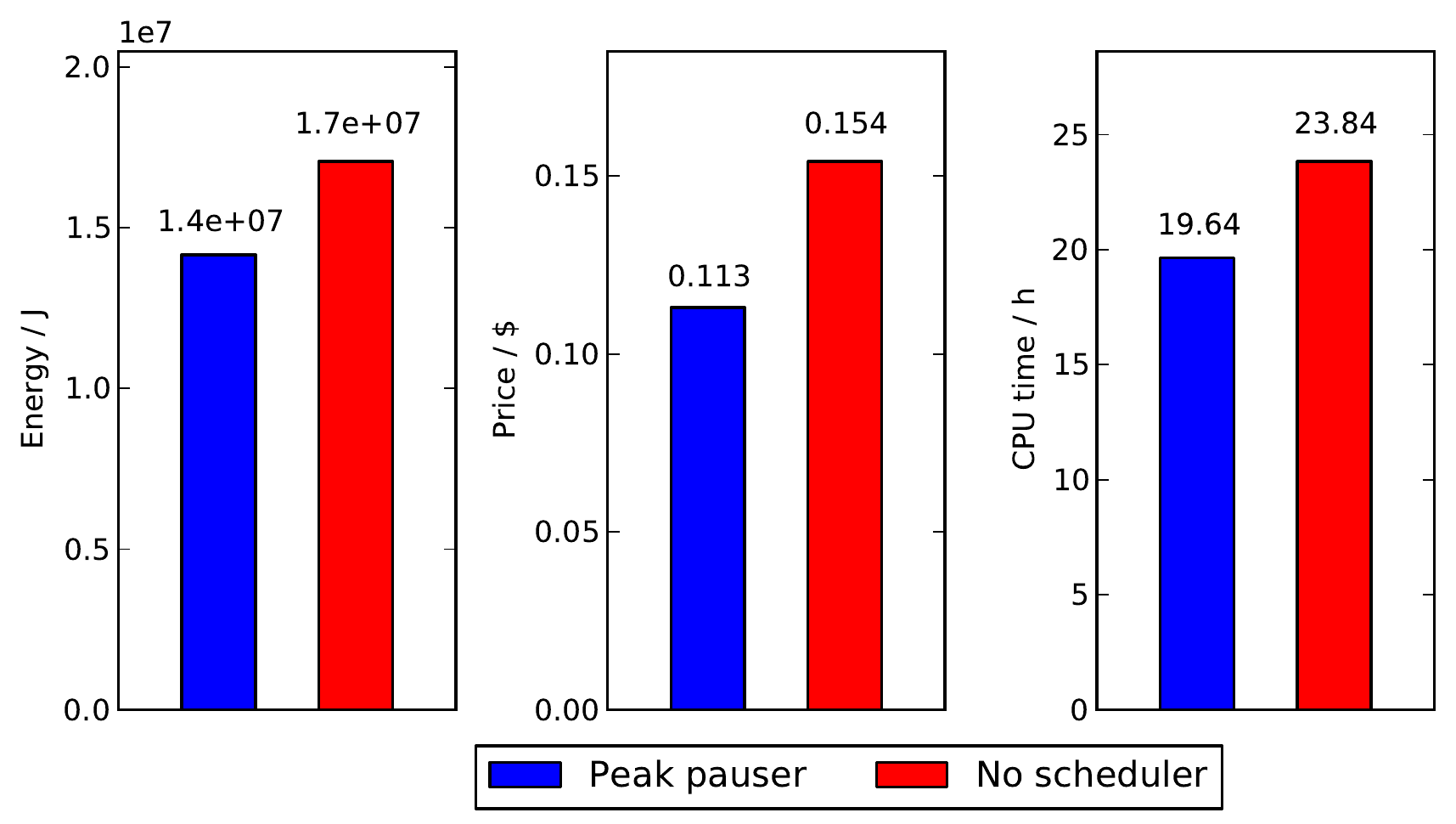}
\caption{Synthetic data results}
\label{fig:aggregated_comparison-synthetic}
\end{figure}

After applying the same energy and price calculation methods on synthetic data\
described in Section \ref{sec:synthetic}, much better savings can be achieved.\
Fig. \ref{fig:aggregated_comparison-synthetic} shows the aggregated results on\
a server with a power of 200 W during peak load and 0 W while being idle. This\
scenario assumes an ideally energy proportional server or the utilization of a\
suspend and wake-on-LAN mechanism on the physical machine (albeit a bit simplified\
as no delay is added to account for these actions).\
The raw energy savings are $\approx 17.1\%$, roughly equal to the\
drop in availability. The price savings exceed both of these values and\
amount to $\approx 26.63\%$ which is a considerable improvement. 

Savings based on more combinations of peak power and idle ratio\
(the ratio of idle and peak power) parameters\
are given in Table \ref{tab:savings}. It is interesting to note how peak power\
does not seem to influence savings much---the difference between 100 and 200 W\
accounts for less than 1\%. The idle ratio, on the other hand, has a high impact,\
underlining the importance of dynamically adjustable, power-proportional servers.


\begin{table}
\centering
\caption{Estimated energy and price savings}
   \begin{tabular}{l||l|l|l|l|}
   \cline{2-5}
&\multicolumn{2}{ c| }{Energy savings} &\multicolumn{2}{ c| }{Price savings}\\\cline{2-5}

	\hline
$idle\ ratio$  |  $P_{peak}$ & 100 W & 200 W & 100 W & 200 W\\
	\hline
	\hline
0\% & 16.96\% & 17.01\% & 26.56\% & 26.63\%\\
30\% & 11.93\% & 11.94\% & 18.68\% & 18.69\%\\
60\% & 6.82\% & 6.82\% & 10.67\% & 10.67\%\\

	\hline
   \end{tabular}
\label{tab:savings}
\end{table}
   	

Based on the estimated annual electricity costs for a company as large as Google\
conservatively placed at \$38M \cite{qureshi_cutting_2009}, even with\
a realistic power usage efficiency of\
1.3 the above savings amass to very large numbers and show considerable impact---both\
economically and environmentally.

\subsection{Resulting \gls{sla}}


Given a 4-hour daily pause, instance availability is 83.3\%.\
If we consider the 0--200 W scenario from\
Fig. \ref{fig:aggregated_comparison-synthetic},\
a \gls{pue} of 1.3 and a \gls{cef} of 1537.82 lb/MWh measured\
in \cite{_united_2008} for Illinois, we can calculate the annual environmental charge-back\
for green instances to be 1300 kg\gls{coe}. This is 300 kg less than a normal instance\
would produce (equivalent to driving an average car for 811 km). This is an approximation,\
but it gives an idea of the order of magnitude and as such might be presented\
to users as an additional advantage of green instances.

If we assume a normal instance cost of \$0.060 per hour,\
the 26.6\% savings in electricity costs would mean that green instances could\
be offered at \$0.044 (disregarding many other factors such as equipment\
amortisation and maintenance costs which are out our our work's scope).

\section{Conclusion}

In this paper we have shown a practical way of utilizing information about\
the electrical grid in the domain of cloud computing\
to visibly reduce energy consumption and costs.

We presented the \emph{peak pauser} scheduling algorithm that offers a clean way of\
managing virtual machines in a computing cloud. It pauses\
computation during hours when the electricity\
price is statistically most probable to peak.\
This mechanism reduces energy costs\
through controllable availability reduction.\
Offering this kind of service as \emph{green instances}\
under special \gls{sla}s to willing\
users only, would ensure that no harm is done from the user point of view.\
Furthermore, this represents environmentally friendly behaviour, because\
energy production applies most stress to the environment\
during times of peak demand,\
when it has to resort to faster and more inefficient generation methods.

Our prototype implementation and experimental evaluation show that\
savings are indeed possible in real-life systems.\
Results stemming from our synthetically-scaled projections of different\
parameters give a sketch of potentially even higher gains if the methods\
were to be used in production systems of today's leading cloud providers.

Overall, maybe the most important aspect of our work is to show the\
importance of collaboration between utility\
companies and cloud providers to implement smarter and more autonomic\
systems that would potentiate sustainable growth of clouds and other\
large-scale computing paradigms.


In the future, we plan to expand this research to other, ``less invasive''\
cloud management actions, such as live migrations. This way,\
the end user would not have to feel any consequences. Also, the\
benchmark procedure should be expanded to cover additional types of\
applications (parallel, memory-, I/O-intensive applications and websites)\
to provide more realistic energy consumption data, assessing a more complete\
scenario.

\textbf{Acknowledgements --} The work described in this paper was\
done through the HALEY project (Holistic Energy\
Efficient Management of Hybrid Clouds), funded by the\
Vienna University of Technology.


\bibliographystyle{IEEEtran-kermit}
\tiny
\bibliography{volatility}

\end{document}